\documentclass[aps,twocolumn,showpacs]{revtex4}
\usepackage{graphicx}
\usepackage[all]{xy}
\newcommand{\be}{\begin{equation}}
\newcommand{\ee}{\end{equation}}
\newcommand{\ben}{\begin{eqnarray}}
\newcommand{\een}{\end{eqnarray}}
\newcommand{\wt}{\widetilde}
\begin{document}

\title{New results for deformed defects}
\author{C.A. Almeida,$^{a,b}$ D. Bazeia,$^a$ L. Losano,$^a$ and J.M.C.
Malbouisson$^c$}
\affiliation{$^a${\small Departamento de
F\'{\i}sica, Universidade Federal da Para\'\i ba, 58051-970,
Jo\~ao Pessoa, PB, Brazil}\\
$^b${Departamento de Matem\'atica, Universidade Regional do Cariri, 63.100-000,
Crato, CE, Brazil}\\
$^c${\small Instituto de F\'\i sica, Universidade Federal da
Bahia, 40210-340, Salvador, BA, Brazil}}

\begin{abstract}
We extend a deformation prescription recently introduced and
present some new soluble nonlinear problems for kinks and lumps.
In particular, we show how to generate models which present the
basic ingredients needed to give rise to dimension bubbles. Also,
we show how to deform models which possess lumplike solutions,
to get to new models that support kinklike solutions.
\end{abstract}


\maketitle

\section{Introduction}

Defects play important role in high energy physics -- see, e.g., Refs.~[1-10]
and references therein. In models described by real scalar fields, defect
solutions are usually topological (kinklike) or non topological (lumplike).
In the present work we deal with models described by a single real scalar
field, and our goal is to extend the deformation procedure introduced in
Ref.~{\cite{blm}} to new models, which support kinklike or lumplike solutions.
To do this, in Sec.~{\ref{sec:dp}} we first consider the stardard procedure.
There we make the deformation prescription as general as possible, and we
introduce new examples. Next, in Sec.~{\ref{sec:ep}} we implement two
distinct extensions, one giving rise to a semi-vacuumless model and the
corresponding domain wall, which serves as seed for generation of dimension
bubbles, as proposed in Refs.~{\cite{BG,mor,GM}}. In the other extension we
show how to implement deformations using non-bijective functions, to deform
models having lumplike solutions to generate new models which support
kinklike solutions. 

\section{Standard procedure}
\label{sec:dp}

We begin with a theory of a single real scalar field in
(1,1) space-time dimensions. The Lagrangian density is usual,
and we use $V=V(\phi)$ to represent the potential which identifies
the model. We also use the metric $(+,-)$, and we work with dimensionless
fields and coordinate. The equation of motion for static fields is
$d^2\phi/dx^2=V^{\prime}(\phi),$ where the prime stands for the
derivative with respect to the argument.
We consider the broad class of potentials having at least one
critical point $\bar\phi$ (that is, $V^{\prime}(\bar\phi)=0$), for
which $V(\bar\phi)=0$. In this case, solutions satisfying the
conditions
\begin{equation}
\lim_{x\rightarrow -\infty} \phi (x) = \bar \phi,
\,\,\,\,\,\,\,
\lim_{x\rightarrow -\infty} \frac{d\phi}{dx} = 0 \, , \label{cond}
\end{equation}
obey the first order equation (a first integral of the equation of motion)
$(d\phi/dx)^2=2 V(\phi(x)).$ For these solutions, the energy densities split
into two equal parts of gradient and potential energy densities.

Many important examples can be presented: the $\phi^4$-model, with
$V_4(\phi)=(1-\phi^2)^2/2$, is the prototype of theories having
topological solitons (kinklike solutions) connecting two minima.
In this case the solutions are $\phi(x)=\pm \tanh(x).$ A situation
where non topological (lumplike) solutions exist is the ``inverted
$\phi^4$-model", with potential given by $V_{4i}(\phi)=\phi^2
(1-\phi^2)/2$. In this case the lumplike defects are
$\phi(x)=\pm\,{\rm sech}(x)$. One notice that the potential
need not be nonnegative for all values of $\phi$ but the solution
must be such that $V(\phi(x))\geq0$ for the whole range
$-\infty < x < +\infty$.

Both topological and non topological solutions can be deformed,
according to the prescription introduced in Ref.~{\cite{blm}}, to generate
infinitely many new soluble problems. This method can be described
in general form via the following statement: Let $f=f(\phi)$ be
a bijective function having continuous non-vanishing derivative.
For each potential $V(\phi)$ bearing solutions satisfying
conditions (\ref{cond}), the $f$-deformed model, defined by
${\wt V}(\phi)= V[f(\phi)]/[f^{\prime}(\phi)]^2,$
possesses solution given by ${\wt\phi}(x)=f^{-1}(\phi(x)),$
where $\phi(x)$ is a solution of the static equation of motion for
the original potential $V(\phi)$.

We prove this assertion by noting that the static equation of motion
of the new theory is written in terms of the old potential as
\begin{equation}
\frac{d^2\phi}{dx^2} =
\frac{1}{f^{\prime}(\phi)}V^{\prime}[f(\phi)] -
2V[f(\phi)]\frac{f^{\prime \prime}(\phi)}{[f^{\prime}(\phi)]^3} \,
. \label{dd1}
\end{equation}
On the other hand, taking the second derivative with respect to
$x$ of the deformed defect ${\wt\phi}(x),$ one finds
\begin{equation}
\frac{d^2{\wt \phi}}{dx^2} = \frac{1}{f^{\prime}({\wt \phi})}
\frac{d^2\phi}{dx^2} - \frac{f^{\prime \prime}({\wt
\phi})}{[f^{\prime}({\wt \phi})]^3} \left( \frac{d\phi}{dx}
\right)^2 \, . \label{dd2}
\end{equation}
It follows from the equation of motion and
from ${\wt\phi}(x)$ that $d^2\phi/dx^2 = V^{\prime}[f({\wt \phi})]$ and
$(d\phi / dx)^2 = 2 V[f({\wt \phi})]$ so that $\wt\phi$ satisfies
(\ref{dd1}), as stated. The ratio between the energy density of the solution
$\phi(x)$ of the undeformed model and the solution
${\wt\phi}(x)$ of the $f$-deformed potential is
${\varepsilon}/{\wt\varepsilon}=({df}/{d\phi})^2.$

Naturally, the deformation procedure heavily depends on the
deformation function $f(\phi )$. Assume that $f:{\bf R}\rightarrow
{\bf R}$ is bijective. In this case,
the $f$-deformation (and the deformation implemented by its
inverse $f^{-1}$) can be applied successively and one can define
equivalence classes of potentials related to each other by
repeated applications of the $f$- (or the $f^{-1}$-) deformation.
Each of such classes possesses an enumerable number of elements
which correspond to smooth deformations of a representative one,
all having the same topological characteristics. The generation
sequence of new theories is depicted in the diagram below.
\begin{widetext}
$$
\xymatrix{
\cdots & \widehat{\widehat{V}} \ar[d]& &\widehat{V}
\ar[d] \ar[ll]_{f^{-1}} & & V
\ar[d] \ar[ll]_{f^{-1}} \ar[rr]^{f}& & \wt{V} \ar[d] \ar[rr]^{f} & &
\wt{\wt{V}} \ar[d] & \cdots\\
\cdots & \widehat{\!\widehat{\phi}}_d & &\widehat{\phi}_d
\ar[ll]^{f}& &\phi_d \ar[ll]^{f} \ar[rr]_{f^{-1}}& & \wt{\phi}_d
\ar[rr]_{f^{-1}}& & \wt{\!\wt{\phi}}_d & \cdots}
$$
\end{widetext}

As an example not considered in Ref.~{\cite{blm}}, take the $\phi^6$-model.
This model, for which the potential $V_6(\phi)=\phi^2(1-\phi^2)^2/2$ has
three degenerated minima at $0$ and $\pm 1$, is important since it
allows the discussion of first-order transitions. It possesses
kinklike solutions, $\phi(x)=\pm\sqrt{\left[1\pm\tanh
(x)\right]/2}$, connecting the central vacuum with the lateral
ones. Take $f(\phi)=\sinh(\phi)$ as the deforming function. The
sinh-deformed $\phi^6$-potential is
\begin{equation}
{\wt V}(\phi) = \frac{1}{2} \tanh^2 (\phi) \left[ 1 -
\sinh^2(\phi) \right]^2 \label{defpfi6}
\end{equation}
and the sinh-deformed defects are
\begin{eqnarray}
{\wt \phi}(x)&=&\pm{\rm arcsinh}\sqrt{\left[1\pm\tanh(x)\right]/2},
\end{eqnarray}
Notice that, since $f^{\prime}(\phi)>1$ for the sinh-deformation,
the energy of the deformed solutions is diminished with respect to
the undeformed kinks. The reverse situation emerges if one takes
the inverse deformation implemented with $f^{-1}(\phi)={\rm
arcsinh}(\phi)$.

Interesting situations arise if one takes polynomial functions
implementing the deformations. Consider $p_{2 n + 1}(\phi)=\sum_{j=0}^{n} c_j 
\phi^{2 j+1}$,
with $c_j > 0$ for all $0 \leq j \leq n$. These are bijective
functions from ${\bf R}$ into ${\bf R}$ possessing positive
derivatives. Fixing $n=0$ corresponds to a trivial rescaling of
the field. For $n=1$, taking $c_0=c_1=1$, one has $f(\phi)=p_3
(\phi)=\phi + \phi^3$ with inverse given by $f^{-1}(\phi)=
(2/\sqrt{3})\sinh[\,{\rm arcsinh}(3\sqrt{3}\phi/2)/3\,]$.
Thus, the $p_3$-deformed $\phi^4$ model, for which the potential
has the form
\begin{equation}
{\wt V}(\phi)=\frac{1}{2} \left(\frac{1-\phi^2-2\phi^4
-\phi^6}{1+3\phi^2}\right)^2 \, ,
\label{p3phi4}
\end{equation}
supports topological solitons given by
\begin{equation}
{\wt \phi}_{\pm}(x)=\pm\frac{2}{\sqrt{3}}\,\sinh\biggl[
\frac{1}{3}\,{\rm arcsinh}\left( \frac{3 \sqrt{3}}{2} \tanh (x)
\right) \biggr]\, .\label{solp3}
\end{equation}
Naturally, the inverse deformation can be implemented leading to
another new soluble problem. But if one takes $n\geq2$, the
inverse of $p_{2n+1}$ cannot be in general expressed
analytically in terms of known functions. This leads to
circumstances where one knows analytically solutions of potentials
which can not be expressed in term of known functions and,
conversely, one has well-established potentials for which
solitonic solutions exist but are not expressible in terms of
known functions.

The procedure can also be applied to potentials presenting non
topological, lumplike, solutions which are of direct interest to
tachyons \cite{tac}. Take, for example, the
Lorentzian lump $\phi_l(x)=1/(x^2 + 1)$
which solves the equation of motion for the potential
$V(\phi)=2(\phi^3 - \phi^4),$ and satisfies conditions (\ref{cond}).
Distinctly of the topological solitons, this kind of solution is not
stable. In fact, the `secondary potential', that appears in the linearized
Schr\"odinger-like equation satisfied by the small perturbations
around $\phi_l(x)$ \cite{Jackiw} is given by
\begin{equation}
U(x) = V^{\prime \prime}(\phi_l (x))=12\frac{x^2 - 1}{(x^2 +
1)^2}\, .
\label{SP}
\end{equation}
This potential is a symmetric volcano-like potential. It has zero
mode given by $\eta_0(x)\sim\phi_l^{\prime}(x)=-2x/(x^2+1)^2$,
which does not correspond to the lowest energy state since
it has a node. Deforming $V(\phi)=2(\phi^3-\phi^4),$  with
$f(\phi)=\sinh(\phi)$ leads to the potential
${\wt V}(\phi)=2\tanh^2(\phi)\left[\sinh(\phi)-\sinh^2 (\phi)\right]$
which possesses the lumplike solution
${\wt\phi}_l(x)={\rm arcsinh[1/(x^2 + 1)]}$.

\section{Extended procedures}
\label{sec:ep}

The deformation prescription is powerful. The conditions under
which our procedure [see Ref.{\cite{blm}}] holds are maintained
if we consider a function for which the contra-domain is an interval of
${\bf R}$, that is, if we take $f:{\bf R} \rightarrow {\rm I} \subset {\bf R}$.
In this case, however, the inverse transformation (engendered by
$f^{-1}:{\rm I} \rightarrow {\bf R}$) can only be applied for
models where the values of $\phi$ are restricted to ${\rm I}
\subset {\bf R}$. We illustrate this possibility by asking for a
deformation that leads to a model of the form needed in Ref.~{\cite{mor}},
described by a ``semi-vacuumless'' potential, in contrast with the
vacuumless potential studied in Ref.~\cite{CV,Bazeia}. Consider
the new deformation function $f(\phi)=1-1/\sinh(e^\phi)$, acting
on the potential $V_4(\phi)=(1-\phi^2)^2/2$. The deformed potential
is
\begin{equation}
\label{dpm} {\wt V}(\phi)=\frac12e^{-2\phi}{\rm sech}^2(e^{\phi})
\left( 2 \sinh(e^{\phi}) - 1\right)^2 \, ,
\end{equation}
which is depicted in Fig.~1. The kinklike solution is
\begin{equation}
{\wt\phi}(x)=\ln\biggl[{\rm
arcsinh}\left(\frac{1}{1-\tanh(x)}\right)\biggr]\, .
\end{equation}
The deformed potential (\ref{dpm}) engenders the required profile:
it has a minimum at $\bar\phi=\ln[{\rm arcsinh}(1/2)]$ and another
one at $\phi\to\infty$. It is similar to the potential required in
Ref.~{\cite{mor}} for the existence of dimension bubbles. The
bubble can be generated from the above (deformed) model, after
removing the degeneracy between $\bar\phi$ and $\phi\to\infty$, in
a way similar to the standard situation, which is usually
implemented with the $\phi^4$ potential, the undeformed potential
that we have used to generate (\ref{dpm}). An issue here is that
such bubble is unstable against collapse, unless a mechanism to
balance the inward pressure due to the surface tension in the
bubble is found. In Ref.~{\cite{mor}}, the mechanism used to
stabilize the bubble requires another scalar field, in a way
similar to the case of non topological solitons previously
proposed in Ref.~{\cite{FGGK}}. This naturally leads to another
scenario, which involves at least two real scalar fields.

\begin{figure}[ht]
\includegraphics[{height=3.5cm,width=8.0cm}]{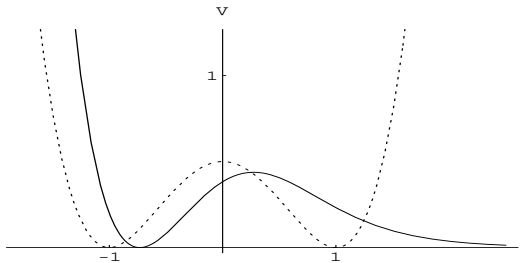}
\caption{The deformed potential ${\wt V}(\phi)$ of
Eq.~(\ref{dpm}), plotted as a function of the scalar field $\phi$;
the dashed line shows the potential of the undeformed $\phi^4$
model.}
\end{figure}

The deformation procedure can be extended even further, by
relaxing the requirement of $f$ being a bijective function, under
certain conditions. Suppose that $f$ is not bijective but it is
such that its inverse $f^{-1}$ (which exists in the context of
binary relations) is a multi-valued function with all branches
defined in the same interval $I\subset {\bf R}$. If the domain of
definition of $f^{-1}$ contains the interval where the values of
the solutions $\phi(x)$ of the original potential vary, then ${\wt
\phi}(x) = f^{-1}(\phi(x))$ are solutions of the new model
obtained by implementing the deformation with $f$. However, one
has to check out whether the deformed potential ${\wt
V}(\phi)=V[f(\phi)]/(f^{\prime}(\phi))^2$ is well defined on the
critical points of $f$. In fact, this does not happen in general
but occurs for some interesting cases.

Consider, for example, the function $f(\phi)=2\phi^2-1$; it
is defined for all values of $\phi$ and its inverse is the double
valued real function $f^{-1}(\phi)=\pm\sqrt{(1+\phi)/2}$, defined
in the interval $[-1,\infty)$. If we deform the $\phi^4$ model
with this function we end up with the potential ${\wt
V}(\phi)=\phi^2(1-\phi^2)^2/2$. The deformed kink solutions are
given by ${\wt\phi}(x)=\pm\sqrt{(1+\phi(x))/2}$ with $\phi(x)$
replaced by the solutions ($\pm\tanh(x)$) of the $\phi^4$ model,
which reproduce the known solutions of the $\phi^6$ theory. The
important aspect, in the present case, is that the $\tanh$-kink
corresponds to field values restricted to the interval $(-1,+1)$
which is contained within the domain of definition of the two
branches of $f^{-1}(\phi)$. The fact that the $\phi^6$ model can
be obtained from the $\phi^4$ potential in this way is
interesting, since these models have distinct characteristics.
Notice that the critical point of $f$ at $\phi=0$ does not disturb
the deformation in this case; this always occur for potentials
having a factor $(1-\phi^2)$, since the denominator of ${\wt
V}(\phi)$ is canceled out. One can go on and
apply this deformation to the $\phi^6$ model; now, one finds the
deformed potential
${\wt V}_6(\phi)=(1/2)\phi^2(1-\phi^2)^2(1-2\phi^2)^2,$
with solutions given by
\begin{equation}
\label{sphi10}{\wt\phi}(x)=
\pm\sqrt{1/2}\sqrt{1\pm\sqrt{\left[1\pm\tanh(x)\right]/2}}\, ,
\end{equation}
corresponding to kinks connecting neighboring minima (located at
$-1$, $-1/\sqrt{2}$, $0$, $1/\sqrt{2}$ and $1$) of the potential,
which is illustrated in Fig.~2. We can repeat the procedure for
the potential ${\wt V}_6(\phi)$, to obtain a sequence of soluble
polynomial potentials, all having exact kinklike solutions. This result
should be contrasted with Ref.~{\cite{lohe}}, which shows that it is
in general hard to find solutions when the model includes higher-order
power in the scalar field.

\begin{figure}[ht]
\includegraphics[{height=3.5cm,width=7.5cm}]{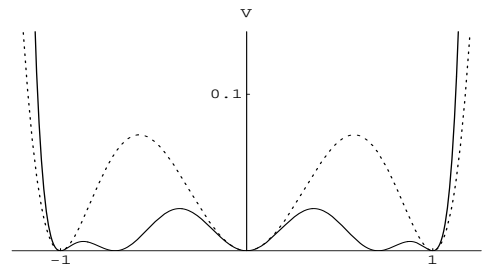}
\caption{The deformed potential ${\wt V}_6(\phi)$ and the undeformed $\phi^6$ 
model (dashed line), plotted as functions of $\phi$.}
\end{figure}

The deformation implemented by the function $f(\phi)=2\phi^2 - 1$
can also be applied to a potential possessing lumplike solutions.
Consider the inverted $\phi^4$ potential
$V_{4i}(\phi)=\phi^2(1-\phi^2)/2$, which has the lump solutions
$\phi(x)=\pm{\rm sech}(x)$. The deformed potential, in this case,
is given by
${\wt V}_{4i}(\phi)=(1/2)(1-\phi^2)\,(\phi^2 - 1/2)^2.$
This potential, which is also unbounded from below, vanishes for
$\phi = \pm1/\sqrt{2}, \,\pm 1$, has an absolute maximum at
$\phi=0$ and local minima and maxima for $\pm1/\sqrt{2}$ and
$\pm\sqrt{5/6}$, respectively. Fig.~3 shows a plot of this
potential. Again, the number of solutions duplicates using such
a deformation: there are two solutions, 
${\wt\phi}(x)_l^{(\pm)}=\pm\sqrt{\left[1+{\rm sech}(x)\right]/2},$
which correspond to lumps running between the local minima
and the lateral zeros of the potential, and also,
\begin{equation}
{\wt\phi}(x)_k^{(\pm)}=
\left\{
\begin{array}{c}
\mp\sqrt{\left[1-{\rm sech}(x)\right]/2} ,\;\;\;\;\;x\leq 0
\\
\pm\sqrt{\left[1-{\rm sech}(x)\right]/2}\;\;\;\;\;\;x\geq 0
\end{array}
\right.
\end{equation}
which correspond to kinklike solutions connecting the minima $\pm1/\sqrt{2}.$
This is a very unique example where non topological or lumplike solutions
are deformed into topological or kinklike solutions.

\begin{figure}[ht]
\includegraphics[{height=4.0cm,width=8.0cm}]{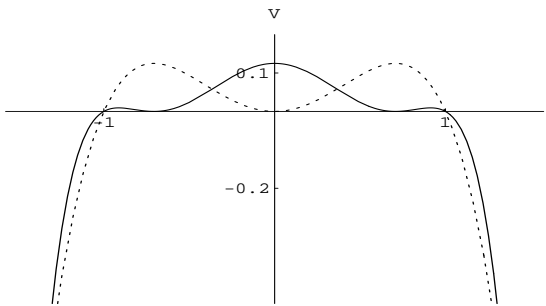}
\caption{The deformed potential ${\wt V}_{i4}(\phi)$ and the
undeformed inverted $\phi^4$ model (dashed line), plotted as functions
of $\phi$.}
\end{figure}

Potentials which have a factor $(1-\phi^2)$ can also be deformed
using the function $f(\phi)=\sin(\phi)$, producing many
interesting situations. In fact, suppose the potential can be
written in the form $V(\phi)=(1 - \phi^2)\,U(\phi).$
This is always possible for all well-behaved potentials that
vanish at both values $\phi=\pm 1$, as shown by Taylor expansion.
Then, the sin-deformation leads to the potential
${\wt V}(\phi)=U[\sin(\phi)],$ which is a periodic potential,
the critical points of $\sin(\phi)$ not causing any problem
to the deformation process. The inverse of the sine function
is the infinitely valued function $f^{-1}(\phi)=
(-1)^k{\rm Arcsin}(\phi)+k\pi$, with $k\in {\bf Z}$ and ${\rm
Arcsin}(\phi)$ being the first determination of $\arcsin(\phi)$
(which varies from $-\pi/2$, for $\phi=-1$, to $+\pi/2$, when
$\phi=+1$), defined in the interval $(-1,+1)$. So to each solution
of the original potential, whose field values range in the
interval $(-1,+1)$, one finds infinitely many solutions of the
deformed, periodic, potential.

Consider firstly the $\phi^4$ model. Applying the sin-deformation
to it, one gets ${\wt V}(\phi)=\cos^2(\phi)/2$ which is one of the
forms of the sine-Gordon potential. The deformed solutions thus
obtained is given by ${\wt \phi}(x)=(-1)^k{\rm
Arcsin}\left[\pm\tanh(x)\right]+k\pi$, which correspond to all the
kink solutions (connecting neighboring minima) of this
sine-Gordon model. For example, the kink solutions $\pm\tanh x$,
which connect the minima $\phi=\pm 1$ of the $\phi^4$ model in
both directions, are deformed into the kinks $\pm{\rm
Arcsin}\left[\tanh (x)\right]=2{\rm Arctan}(e^{\pm x}) - \pi /2$
(which runs between $-\pi /2$ and $\pi /2$) if one takes $k=0$
while, for $k=1$, the resulting solutions connect the minima $\pi
/2$ and $3\pi /2$ of the deformed potential.

This example can be readily extended to other polynomial
potentials, leading to a large class of sine-Gordon type of
potentials. For instance, the $\phi^6$ model,
$V(\phi)=\phi^2(1-\phi^2)^2/2$, deformed by the sine function,
becomes the potential ${\wt V}(\phi)=(1/2)\cos^2(\phi)[1 -\cos^2(\phi)],$
which has kinklike solutions given by
\begin{equation}
{\wt \phi}(x)=\pm (-1)^k{\rm Arcsin}\sqrt{\left[1
\pm\tanh(x)\right]/2}+k\pi .
\end{equation}
On the other hand, if one considers $V(\phi)=(1-\phi^2)^3/2$,
which is unbounded below and supports kinklike
solutions connecting the two inflection points at $\pm 1$, one
gets the potential ${\wt V}(\phi)=(1/2)\cos^4(\phi),$ which
is solved by ${\wt\phi}(x)=\pm (-1)^k{\rm Arcsin}(x/\sqrt{1+x^2})
+ k\pi .$

Another particularly interesting situation where non topological
solutions are deformed into topological solutions appears if one
consider the inverted $\phi^4$ model, which presents lumplike solutions.
The sin-deformation of the potential $V(\phi)=\phi^2(1 - \phi^2)/2$
leads to the potential ${\wt V}(\phi)=\sin^2(\phi)/2$. In this
case, the lump solutions of $V(\phi)$, namely $\phi(x)=\pm {\rm
sech}(x)$, are deformed into ${\wt \phi}(x)=\pm (-1)^k{\rm
Arcsin}\left[{\rm sech}(x)\right]+k\pi$. Consider the
$(+)$-solution and take initially $k=0$. As $x$ varies from
$-\infty$ to $0$, ${\rm sech}(x)$ goes from $0$ to $1$, and ${\rm
Arcsin}\left[{\rm sech}(x)\right]=2{\rm Arctan}(e^x)$ changes from
$0$ to $\pi /2$. If one continuously makes $x$ goes from $0$ to
$+\infty$, then the deformed solution passes to the $k=1$ branch
of $\arcsin(\phi)$, $-{\rm Arcsin}\left[{\rm sech}(x)\right]+\pi$
($=2{\rm Arctan}(e^x)$ for $0\leq x < +\infty$), which varies from
$\pi /2$ to $\pi$ as $x$ goes from $0$ to $+\infty$. Thus, in this
case, the lump solution $+{\rm sech}(x)$ of the inverted $\phi^4$
model is deformed in the kink of the sine-Gordon model connecting
the minima $\phi=0$ and $\phi=\pi$. Under reversed conditions
(taking the $k=1$ branch before the $k=0$ one), the lump solution
$-{\rm sech}(x)$ leads to the anti-kink solution of the
sine-Gordon model running from the minimum $\phi=\pi$ to $0$. The
other topological solutions of the sine-Gordon model are obtained
considering the other adjacent branches of $\arcsin(\phi)$.

\section{Comments and conclusions}
\label{sec:comments}

In the former work on deformed defects \cite{blm} we have stressed that
the deformation procedure strongly depends on a function $f=f(\phi),$
the deformation function, and there we have only considered bijective
functions that obey $f:{\bf R}\to{\bf R}.$ In the present work,
we have extended
the deformation procedure with the inclusion of two new possibilities.
First, we have considered deformation functions such that
$f:{\bf R}\to{\bf I}$ with ${\bf I}\subset {\bf R},$ which gives rise
to new models such as the one
recently considered in Ref.~{\cite{mor}}, which requires a semi-vacuumless
potential. Furthermore, we have shown how to deal with non-bijective
functions to build new models. This last case leads to very interesting
possibilities of deforming models which support non topological defects,
to give rise to models which support topological defects.

We would like to thank W. Freire for comments, and CAPES,
CNPq, PROCAD and PRONEX for financial support. CAA thanks FUNCAP
for a fellowship.


\begin{thebibliography}{99}
\bibitem{blm}D. Bazeia, L. Losano and J.M.C. Malbouisson,
Phys. Rev. D {\bf66}, 101701(R) (2002).
\bibitem{BG}S.K. Blau and E.I. Guendelman, Phys. Rev. D
{\bf40,} 1909 (1989).
\bibitem{mor}J.R. Morris, Phys. Rev. D {\bf67,} 025005 (2003).
\bibitem{GM}E.I. Guendelman and J.R. Morris, Phys. Rev. D
{\bf68}, 045008 (2003).
\bibitem{tac}J.A. Minahan and B. Zwiebach, J. High Energy
Phys. {\bf09,} 029 (2000).
\bibitem{Jackiw}R. Jackiw,  Rev. Mod. Phys. {\bf49,} 681 (1977).
\bibitem{CV}I. Cho and A. Vilenkin, Phys. Rev. D {\bf59,}
021701(R) (1999); {\bf59,} 063510 (1999).
\bibitem{Bazeia}D. Bazeia, Phys. Rev. D {\bf60,} 067705 (1999).
\bibitem{FGGK}J.A. Frieman, G.B. Gelmini, M.Gleiser, and
E.W. Kolb, Phys. Rev. Lett. {\bf60,} 2101 (1988).
\bibitem{lohe}M.A. Lohe, Phys. Rev. D {\bf20,} 3120 (1979).
\end{thebibliography}
\end{document}